# SecureTrack- A contact tracing IoT platform for monitoring infectious diseases

**Authors:**
**Shobhit Aggarwal**
University of North Carolina at Charlotte,
Email: shobhuagg@gmail.com

**Arnab Purkayastha**
Western New England University
Email: arnab.purkayastha@wne.edu

___________________________________________________________________

**Abstract**

The COVID-19 pandemic has highlighted the need for innovative solutions to monitor and control the spread of infectious diseases. With the potential for future pandemics and the risk of outbreaks particularly in academic institutions, there is a pressing need for effective approaches to monitor and manage such diseases.

Contact tracing using Global Positioning Systems (GPS) has been found to be the most prevalent method to detect and tackle the extent of outbreaks during the pandemic. However, these services suffer from the inherent problems of infringement of data privacy that creates hindrance in adoption of the technology.

Non-cellular wireless technologies on the other hand are well-suited to provide secure contact tracing methods. Such approaches integrated with the Internet of Things (IoT) have a great potential to aid in the fight against any type of infectious diseases.

In response, we present a unique approach that utilizes an IoT based generic framework to identify individuals who may have been exposed to the virus, using contact tracing methods, without compromising the privacy aspect. We develop the architecture of our platform, including both the frontend and backend components, and demonstrate its effectiveness in identifying potential COVID-19 exposures (as a test case) through a proof-of-concept implementation. We also implement and verify a prototype of the device. Our framework is easily deployable and can be scaled up as needed with the existing infrastructure.

1. **Introduction**

Infectious diseases pose a significant threat to public health and can have devastating consequences. Traditional methods of contact tracing, which rely on human reporting and memory, can be slow and prone to errors. Moreover, GPS based contact tracing systems cause data privacy infringement issues. With the advent of the Internet of Things (IoT), there is an opportunity to develop more advanced and efficient contact tracing systems. This article explores a contact tracing IoT platform that leverages wearable sensors and beacon based tracker to monitor individuals' movements and interactions. At the same time, all the tracking information is encrypted and stored locally on the device. This not only removes privacy concerns but also gets rid of user involvement in the storage and retrieval of data. For the rest of the article, we make use of the ongoing COVID19 pandemic as a representative scenario for illustrating the effect of highly contagious diseases.

The SARS-CoV-2 or the novel coronavirus(COVID-19) pandemic has spread across more than 220 countries with a total of 676,609,955 confirmed cases resulting in the death of 6,881,955 lives at the time of this writing [1]. Governments and medical communities all across the world are putting up a wartime effort to mitigate this major crisis. There has been research on development of single or combination of various measures that includes vaccines and therapeutic agents [2, 3, 4].

Table 1 reports the COVID-19 data in the United States since the beginning of the pandemic [5] [6]. We specifically look at the young working population in the age group of 0-64 yrs. This particular age group that belongs to children, students and working adults have shown evidential amounts of resilience against the COVID-19 pandemic. Much of this can be attributed to the enforced lockdown measures adopted by governments across the world [7, 8]. However, long periods of shutdown has been found to be detrimental to both the mental and physical health of such individuals [8, 9, 10, 11, 12, 13, 14, 15]. This leads to the need of reopening of academic institutions at an early means without risking lives.

| | |
|---|---|
| **Total cases** | 103 Mn+ |
| **Total deaths** | 1.1 Mn+ |
| **Total deaths age group (0-69)** | 20.2% of total deaths |

Table 1: COVID-19 Data United States

Unfortunately, this age-group has been found to be the largest asymptomatic carriers (40% people) of the virus [16]. School going children in particular are the most susceptible candidates since there is a large possibility that children become carriers and help in accelerating the spread of the virus once they come in contact with their families. The same is true for university and college students and academicians. Moreover, the healthcare burden [17] involved with such a large group is simply enormous. This makes reopening a long and distant dream without the immediate availability of vaccines.

*1.1. Reopening academic institutions*

Reopening of academic institutions has been a serious topic of debate across the spectrum when it comes to 'safe and phased' reopening guidelines issued by the Centre of Disease Control, US (CDC) [18]. Shutdown of schools and academic institutions have resulted in a lot of hassles for students and children in particular. While online teaching is the safest go-to option during these times, lack of physical presence of teachers and classroom training is an inefficient solution.

There have been many works that discuss safe reopening of academic institutions [19, 20]. These works focus mainly on frequent screening of students for COVID-19 symptoms. While increased screening is definitely a good and safe way to detect an outbreak, however it does not account for asymptomatic individuals who may inadvertently spread the disease among peers and family members. Moreover, rapid high-specificity tests may not be cost-effective to deal and control outbreaks of the virus. This is evident from the phased openings of schools in several parts of the country that have resulted in aggravations [5] that led to immediate closure of all schools.

*1.2. Contact Tracing for COVID-19*

Contact Tracing [21] has effectively proven to be an important mechanism to identify and deter the spread of COVID-19. Various approaches are being adopted by governments and authorities across the world [22, 23, 24] to locate and diagnose vulnerable patients who have underlying conditions. A majority of the contact tracing solutions are in the form of smartphone based apps [22, 24, 25] that require the user to install an iOS or Android version of the app [25] on their phone, while the app keeps a track of the mobile data of users. These apps either use the built-in 'GPS' viz. location services of the smartphone [25] or ask the user to physically update a patient's health condition thereby updating the COVID-19 database in that region. This data is then shared with the medical authorities and necessary steps are taken to mitigate the problem as soon as an individual is infected. While the initial response of the general public for adoption of these apps have largely been positive, recently a steady decline in app downloads was observed [25, 26].

Contact tracing applications suffer from numerous drawbacks that can be categorized in the following areas.

1. Infrastructure- The availability of smartphones and uninterrupted network connectivity is an important factor when considering the success of mobile-based apps. This problem is more pronounced in developing and under-developed countries where the majority of people do not have access to the required infrastructure. Moreover, certain groups of individuals like the children and elderly, who are particularly susceptible to an infection may not have the expertise to use a smartphone by themselves.

2. Data Privacy- Data privacy is perhaps the most serious concern in the use of mobile-based contact tracing apps. An eminent vulnerability that arises from massive sharing and storage of user data is in the form of surveillance that has the possibility to be used and exploited by agencies or corporations handling the data.

3. Incorrect information- Physically entering correct information is another key factor in these apps which put complete responsibility on the user. However, in communities [27, 28, 29] where infectious diseases can be a social stigma, people are generally afraid to accept their illness and will refrain from accurately updating their health condition. Further, there is a huge possibility of spread of fake news to sensationalize the situation through the use of these apps.

In addition to this, contact tracing apps are practically of no use when it comes to asymptomatic individuals who can roam freely undetected and cause sufficient damage to healthy individuals.

*1.3. Wearables in IoT*

IoT platforms have reached consumers nowadays with smart and easily accessible wearable devices [30, 31]. Wearable technology is not only prevalent in the fitness market in the form of trackers and smartwatches but is also growing rapidly in the healthcare industry for monitoring patients both on and off hospitals [32, 33, 34]. The seamless integration of edge and cloud based infrastructure has helped wearables these days to find numerous applications as ecosystems of connected devices [35]. Moreover, wearable devices have tremendous potential benefits [36, 37] for healthcare research with the ability to reach more and more users and guide healthcare professionals to provide better diagnosis. Figure 1 shows the entire spectrum of wearable devices that play an important role in our lives [38].

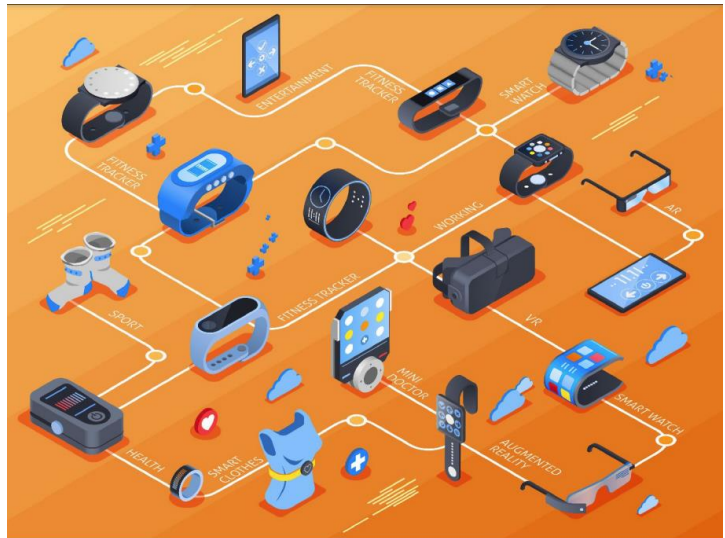
Figure 1: Wearables in IoT

One of the major drawbacks of wearables is the issue of feasibility. Most users do not like to get into the nitty grity of connection, software download and active data input to these devices which can be a headache sometimes. This is one key reason that wearable technology hasn't yet found a sizeable chunk of the IoT device market.

Our wearable tracker allows monitoring of users for COVID-19 positive symptoms without the hassle of user involvement in the storage and retrieval of data.

*1.4. SecureTrack*

SecureTrack is a contact tracking device that has the potential to autonomously keep track of all the people with whom a person comes in contact with. The device does not need any location service to function. All the information is encrypted and is stored locally on the device. Only an authorized official has access to the information stored on the device.

The device can be embedded in a face mask or flexibly worn as a wrist band, a necklace pendant or can be kept in the pocket.

*1.5. Contributions*

In summary, the contributions of this paper are as follows :-

- Propose a generic framework based on contact tracing solutions for fast, safe and efficient reopening of academic institutions.

- Introduce a novel wearable device for early and quick detection of COVID-19 without making user data vulnerable to privacy breach.

- Develop a prototype of SecureTrack framework using MICAz as the target platform as proof-of-concept.

- Cost analysis for mass-level production and distribution of our device.

We have open-sourced all the source codes of SecureTrack here- https://github.com/arnabapurk/SecureTrack.

The next section presents works done in the related domain and the challenges associated.

## 2. Related Works

IoT technology in medical applications have come a long way due to the advent of edge and cloud based integrated platforms. Home Health Hub Internet of Things [39] for continuous monitoring of elderly patients is a framework structured into layers that sense various physiological activities by different sensors such as the electrocardiogram (ECG), pulse oximeter, electromyogram (EMG), respiratory meter, blood glucose sensor, and blood pressure sensor. The information related to the user's health condition is immediately received by the user application layer. Mobility, delay tolerance, low set up cost, user-friendliness, clear layered design are some of the advantages of this framework.

A mobile application based platform is presented in [40]. The application collects ECG data using bio-sensors and uploaded the data to the cloud and creates a medical record for each patient. This framework claims to potentially decrease the waiting time in hospitals. The model can also minimize personnel and administrative costs.

[41] presents a health monitoring system called iHome Health IoT. This system connects the individual home environment with hospital, emergency center, and other medical facilities and provides remote prescription and medication non-compliance services. Parameters such as ECG signals are measured and if misuse of medication is detected, it alerts caregivers.

[42] presents a health monitoring system based on LoRaWAN. The main objective of this work is to provide a secure and low-cost health monitoring system. Their end-node is composed of ehealth sensors that consists of glucometer, a blood pressure sensor

and a thermometer. This system was able to monitor chronic diseases such as diabetes and arterial hypertension.

Contact tracing for detection and tracing of COVID-19 patients is an ongoing topic of interest. Current use of IoT for contact tracing is mainly in the form of smartphone based apps that rely heavily on user data of their respective locations and individual user input. Adoption and use of such applications have many significant drawbacks as discussed in depth in Sec 1.

This work proposes a unique solution that is secure with respect to data privacy, inexpensive and readily usable. Our device can be easily made available and deployed across clusters of large institutions like schools and colleges thus aiding effective contact tracing to aid in quick reopening of academic institutions. The next sections present a detailed description of the proposed framework.

## 3. Methodology

This section describes the workflow of the SecureTrack device. The device mainly consists of a radio module, memory module, processing unit, and battery module.

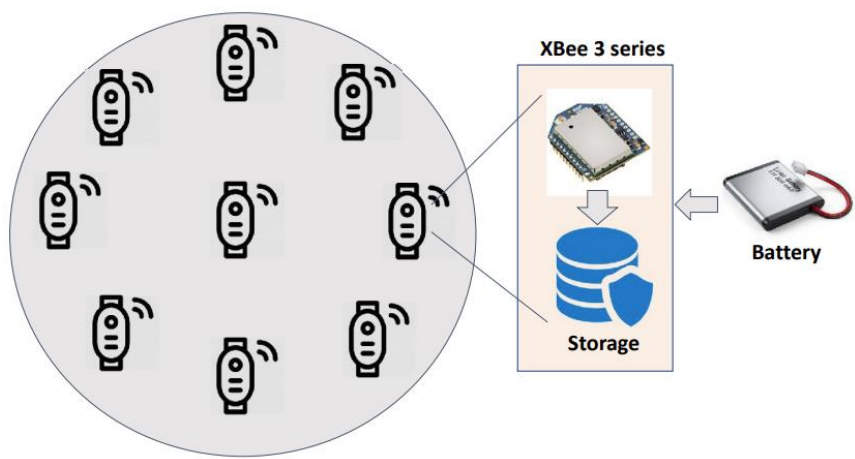

Figure 2: Basic components of SecureTrack

*3.1. Modules*

The SecureTrack device uses XBee 3 series [43] module as a radio unit. XBee is a Zigbee based device that uses IEEE 802.15.4 specifications. The main reason to select Zigbee over other technologies such as ANT+ and Bluetooth is that in ANT+ the devices are configured in a master-slave configuration. The devices configured as slave are not allowed to communicate with other devices directly. The communication has to go through the master node. Although, Bluetooth devices can broadcast beacons with the advent of Bluetooth Low Energy (BLE), such devices are only transmitters not receivers. XBee devices are very cost-effective and consume very less power. The transmission range of these devices can be easily controlled by changing the transmit power [44]. It also has the required computational capabilities thus eliminating the need for an external microcontroller. The device can be easily programmed using a micropython interface. On-chip flash memory is used to store the information and a rechargeable Li-Po battery is used to power the device. Figure 2 shows the basic components of the SecureTrack device.

SecureTrack periodically broadcasts a transmission with the transmit power high enough that the communication can reach a range of 6 feet. After airing, the device enters into listen mode and starts listening to the broadcasts made by other devices within 6 feet radius. The broadcast message only contains the device ID that is unique to every SecureTrack device. All the received messages are stored in the memory module of the device.

*3.2. Architectural framework*

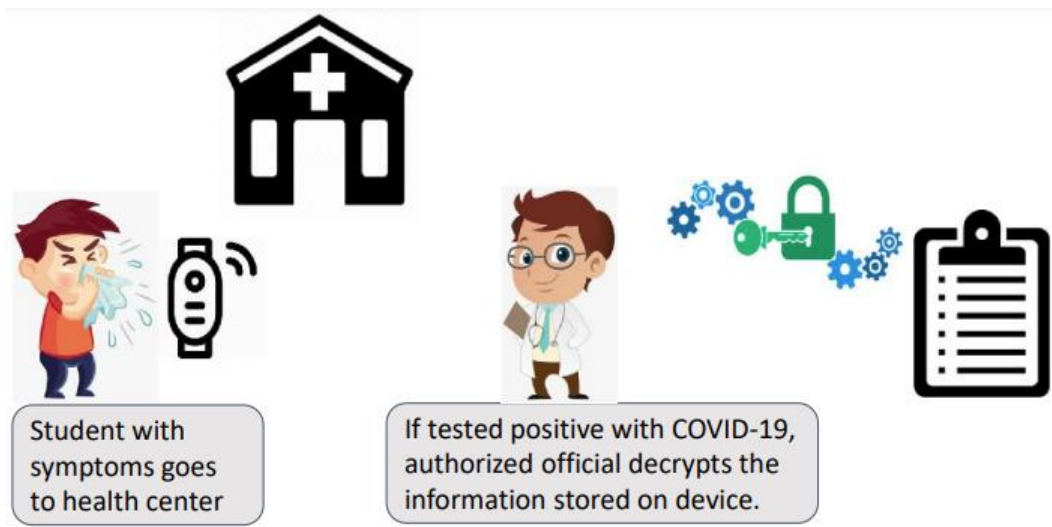

Figure 3: Framework of the device

Figure 3 shows the conceptual framework of our device. Here we assume that every school/college in any county across the nation has access to at least one student health center or medical provider facility that can keep records of the student COVID-19 data.

Whenever someone is tested positive for COVID-19, his/hers' SecureTrack device will be accessed by an authorized official. This official has the required resources to generate the decryption key. The device IDs will be mapped to the individual student IDs. Once the stored data is decrypted, the database of all those students who have come in contact with the affected student in the past 14 day period will be available securely to the student health center. A notification (mail/phone) can then be instantly sent to the affected students/ guardians to get themselves tested for COVID-19. From the current studies [16, 19] we can safely say that only 60% of all such individuals will be symptomatic carriers, the rest will be asymptomatic.

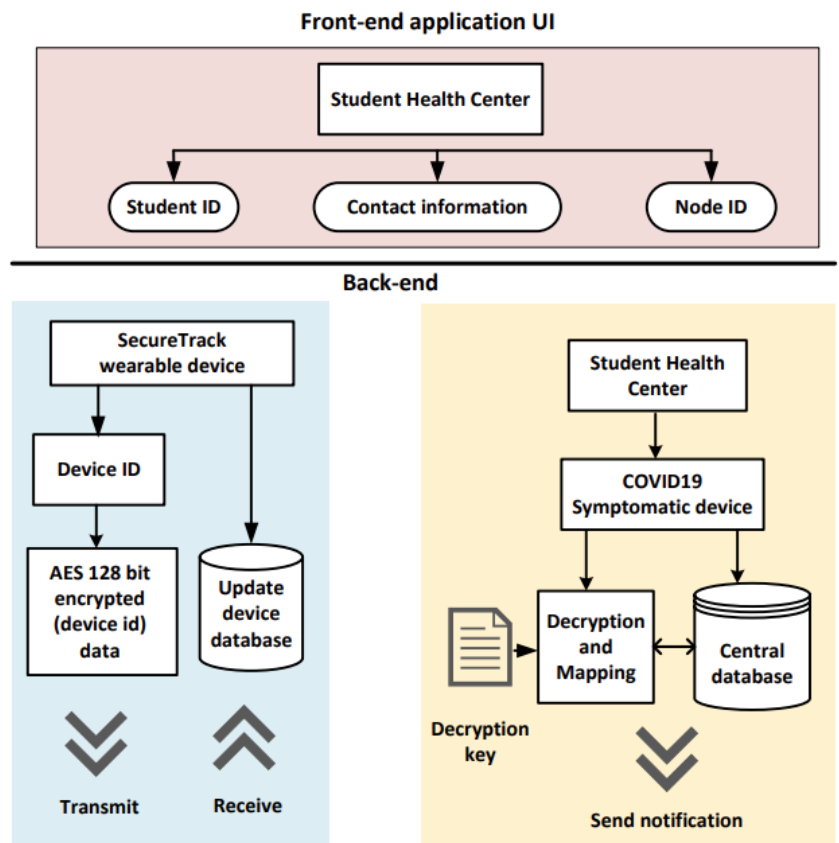

Figure 4:Detailed architecture of SecureTrack

Next, we look at the detailed architecture of SecureTrack (Fig 4). We have divided the work-flow in 2 sections, namely, *Front-end* and *Back-end*.

**Front-end application UI**- At the front-end is our web-based application user interface (UI) that is accessible to the authorized health official of the respective student health center. The dashboard has only 3 tabs viz., Student ID, Contact information (email address or phone number where the individual wants the information to be sent) and device generated Node ID. Whenever a symptomatic individual comes for COVID testing, this information will be registered in the Central database.

**Back-end**- The SecureTrack back-end comprises 2 parts. In the first part (highlighted in blue), the device ID is encrypted using the AES 128 bit encryption and periodically broadcasted. Simultaneously, periodic broadcasts from nearby devices are received and the device memory is filled with all such encrypted device id data. The working of the device is further explained in Sec 3.3. In the second part (highlighted in yellow), whenever an individual is tested COVID-19 positive, the individual's device is identified as 'COVID-19 symptomatic device'. Next, a decryption key is generated from a combination of the Student ID and the Node ID. This key is then used to decrypt the encrypted data in the devices' database that is mapped to identify all other devices that have come in contact with the individual in the past 14 days. The mapping is done by the central database. Once mapping is processed, the individual's who have come in contact can be notified.

In this way, each time a new case is registered in the health center, all possible COVID-19 positive cases in the individual's contact can be easily identified and quarantined, thus breaking the chain of uncontrolled transmission in the form of both symptomatic/ asymptomatic carriers.

*3.3. Working*

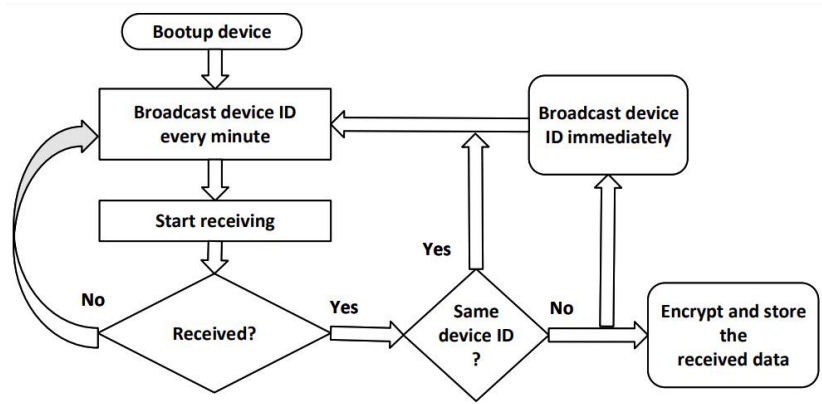

Figure 5: Flowchart of the design

Figure 5 explains the detailed working of the SecureTrack device. Once the device boots up, it will periodically broadcast its device ID every minute. After broadcasting, the device will open a receive window.

If the device receives a transmission made by some other device, it will perform a check on the received device ID. If this device ID is not received within the last 24 hours it will perform two tasks simultaneously. Firstly, it will encrypt the received message using an AES-128 bit encryption. The encryption key is unique to every device and will be generated from the device ID itself. Secondly, it will broadcast its device ID once in addition to the periodic broadcast. This broadcast is required to make sure that the periodic transmission has not missed the receive window of any other device in the vicinity. The encrypted data will be stored on the device for 14 days, and the redundant device IDs will be replaced with the most updated reception.

If the device ID received is the same as the device ID received within the time frame of 24 hours, then the device will consider this reception redundant and will return to the periodic one minute broadcast to avoid getting stuck in a broadcast-reception loop and drain the resources unnecessarily.

Equation 1 is used to calculate the memory requirements for the device.

$$Memory(KB) = \sum_{i=1}^{i=14} \frac{S_i * 128}{8 * 1024} \qquad (1)$$

Where,
*i* is the day number ranging from 0 through 14.
$S_i$ is the number of contacts a student has made with on the '*i*' th day.

Assuming a class size of 50 students for illustrative purposes. Supposing on an average a student comes in contact with 100 other students. As we are saving the data entries for a period of 14 days and all the data is encrypted using AES(128), hence each entry will be 128 bit long. Putting this in equation 1 we get a memory requirement of 21.88 KB. Since the students come in contact with their classmates on a daily basis, there will be a lot of redundant data. Hence, the actual memory requirement will be much lower than calculated. XBee 3 series has an on-board memory of 128KB ROM [43] that is more than sufficient to handle the amount of data collected.

## 3.4. Cost Analysis

In this section we do a detailed cost analysis study of our proposed device. The price of the device can be easily determined based on its individual components that are the XBee module and the battery. We have tried to make our device as bare-bones as possible so as to be cheaply available and readily scalable.

As an illustration we compare our proposed device with a couple of cheap and popular fitness trackers available in the market. We choose Honor band 5 fitness tracker [45] and Xiaomi Mi Band 4 fitness tracker [46] for this comparison. Table 2 shows the specification and the features of the SecureTrack device in comparison to these fitness trackers. Here we can see that since we employ minimalistic features for our device, the price associated with it can be several orders of magnitude cheaper than the commercial bands. We are sure that in a fabrication facility with bulk manufacturing even this cost can be reduced to a minimum.

| Features | HONOR Band 5 | Xiaomi Band 4 | SecureTrack |
|---|---|---|---|
| Display Type | AMOLED 0.95" | AMOLED 0.95" | X |
| Touchscreen type | On-cell capacitive touchscreen | On-cell capacitive touchscreen | X |
| Screen protection | 2.5D tempered glass | 2.5D tempered glass with anti-fingerprint coating | X |
| Button | Single touch button | Single touch button | X |
| 3-axis accelerometer | ✓ | ✓ | X |
| 3-axis gyroscope | ✓ | ✓ | X |
| Heart rate sensor | ✓ | ✓ | X |
| SpO2 Monitor | ✓ | X | X |
| Vibration Motor Type | Rotor | Rotor | X |
| Water Resistant | ✓ | ✓ | X |
| Wireless Module | Bluetooth | Bluetooth | Zigbee |
| On-Chip RAM | - | 512 KB | 128 KB |
| On-Chip ROM | - | 16 MB | 1 MB |
| Remote Picture Taking | ✓ | ✓ | X |
| Battery | 135 mAH | 100 mAH | 100mAh |
| Price | $31.99 | $42.62 | - |

Table 2: Cost Comparison analysis

*3.5. Limitations*

Few of the limitations of the framework include:

- The academic institutions need to make the device compulsory for all students to wear at all times. However, due to the low cost infrastructure involved, the device can be made freely available to the students.

- The battery life of our SecureTrack is still a matter of speculation since we have not yet fabricated a physical device. However, both the illustrative devices in Table 2 incorporate a 135 mAh and 100mAh battery that run for 20 days and 14 days on full charge respectively [45, 46]. Our bare-bones device should theoretically run for a lot more than that if not for less. In any case, the user is expected to make sure to charge the device as needed.

The next section discusses an implementation and results of the proposed framework on a demonstrative IEEE 802.15.4 based hardware.

### 4. Implementation

We have implemented a base-level model of our framework to validate the correct functioning of our algorithm in this section. We further share our initial results on demonstrative hardware. Due to the unavailability of the proposed hardware, we employ CrossBow(XBOW) MICAz devices [47] for experimental verification purposes. Both MICAz and XBee operates on the IEEE 802.15.4 standard, hence we chose the same for prototyping our device.

For debugging and demonstration purposes we have made the following assumptions:

- Received Signal Strength Indication (RSSI) is used to calibrate the distance for classroom environment. The device can be adapted to various environmental settings by further calibrating the RSSI's according to the Path Loss exponent [48].

- The Transmit power is not restricted to calibrate the RSSI of the device that is used to measure the 6 feet radius. This assumption can be easily taken care of as discussed in Sec. 3.1.

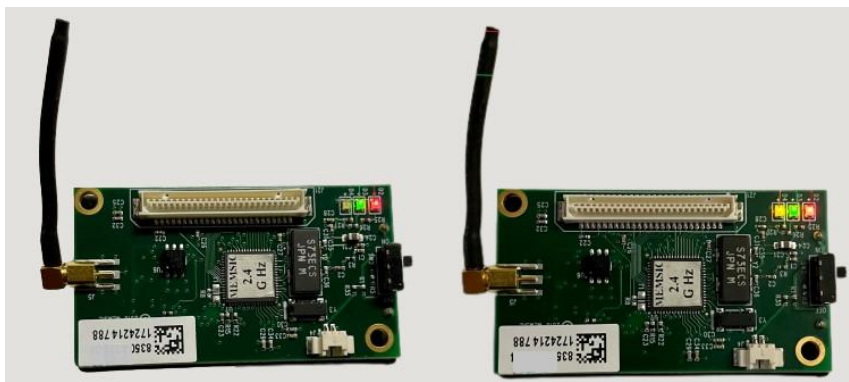

Figure 6: LED indicators of MICAz for SecureTrack. *Red*- Transmitting, *Green*- Receiving, *Yellow*-Unsafe distance.

As shown in Fig 6, MICAz has three onboard LED's that are used to indicate various states viz., toggling of Red LED indicates Beacon transmission, toggling of Green LED indicates Beacon reception and setting of Yellow LED indicates that any other node is within 6 feet vicinity of the current node.

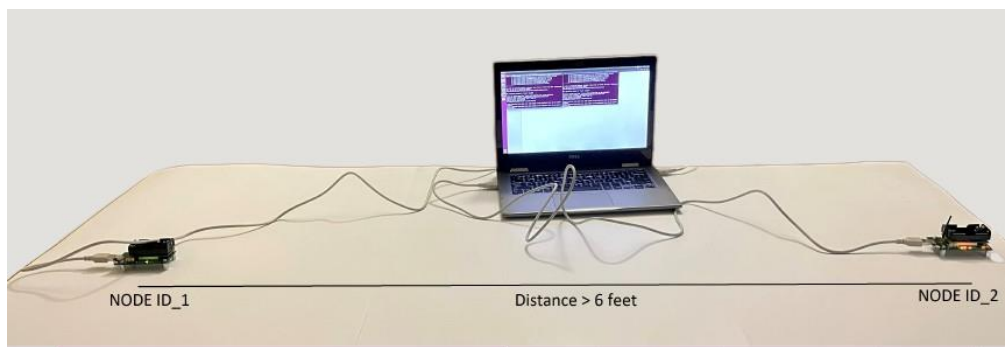

Figure 7: Setup of two nodes more than 6 feet apart

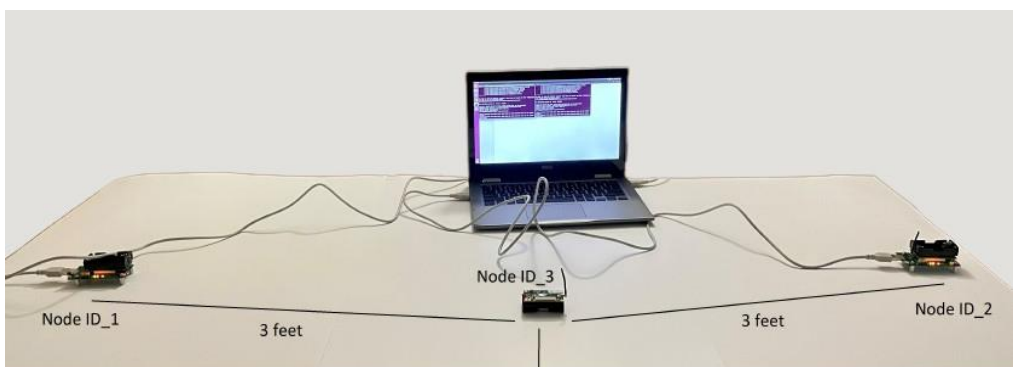

Figure 8: Setup of three nodes when two nodes(ID_1 and ID_2) are more than 6 feet apart and Node ID_3 is in between.

Figure 7 shows two MICAz devices with 'Node ID_1' and 'Node ID_2' separated by a distance greater than 6 feet connected serially with a laptop that acts as a server to display the serial output of the communicating devices on the terminal. Since the transmit power is not restricted the beacon transmission will take but as the distance is more than 6 feet, the node IDs will not be saved in the device database.

Figure 8 has the same setup as Figure 7 but with a third device ('Node ID_3') placed equidistant from the 2 devices at 3 feet, thus violating the 6 feet threshold. Due to this, nodes with IDs 1 and 2 will save the 'Node ID_3' in their database while device with 'Node ID_3' stores both the node IDs 1 and 2.

| DUT | Received Beacon Device_ID | RSSI | Distance (feet) | Action |
|---|---|---|---|---|
| Node ID_1 | Node ID_3 | -30 | >6 | Discarded |
| Node ID_1 | Node ID_4 | -18 | <6 | Entry Saved |
| Node ID_1 | Node ID_7 | -24 | <6 | Entry Saved |
| Node ID_1 | Node ID_2 | -21 | <6 | Entry Saved |
| Node ID_1 | Node ID_3 | -18 | <6 | Entry Saved |
| Node ID_1 | Node ID_4 | -32 | >6 | Discarded |
| … | … | … | … | … |
| Node ID_7 | Node ID_5 | -15 | <6 | Entry Saved |
| Node ID_7 | Node ID_1 | -24 | <6 | Entry Saved |
| Node ID_7 | Node ID_3 | -12 | <6 | Entry Saved |
| Node ID_7 | Node ID_4 | -23 | <6 | Entry Saved |
| Node ID_7 | Node ID_3 | -30 | >6 | Discarded |
| … | … | … | … | … |

Table 3: Implementation results on DUT and Received Beacon Device_ID

To verify our implementation, we have scaled this to seven MICAz devices. In this case, instead of laying the devices on a flat platform, they were handheld to accurately simulate a classroom environment. Table 3 presents a snippet of results obtained on 2 nodes, 'Node ID_1' and 'Node ID_7'. For each case, the 'Device-Under-test (DUT)' is fixed while the rest of the devices are allowed to move. The RSSI value is pre-calibrated to -25 to measure a distance of 6 feet between the devices. Following conditions are tested, based off which an action is performed:-

- If  RSSI > -25 , indicates that the distance between the 'Device-Under-test (DUT)' and 'Received Beacon Device\_ID' is less than 6 feets. The corresponding Node ID entry is saved on both the devices.

- If RSSI < -25 , indicates that the distance between the 'Device-Under-test (DUT)' and 'Received Beacon Device\_ID' is more than 6 feets. The corresponding Node ID entry is discarded on both the devices.

- Finally, if a beacon is received from the same device, the current RSSI value of that device is checked and the corresponding entry is updated (either saved or discarded) on both the devices

### 5. Future

The COVID 19 pandemic has shown the importance of contact tracing for any communicable disease. The SecureTrack device can play a pivotal role in tracing, tracking and containing the spread of any communicable disease or pandemics. The proposed framework is easy to scale up across industries, commercial complexes and healthcare facilities.

In order to be deployed in such large clusters, the student ID can be replaced with other preexisting national or state IDs of citizens like Driving License, SSN etc. This data can be securely encrypted and managed by a common center for medical records at local level for effective contact tracing.

### 6. Conclusions

The novel coronavirus (COVID-19) pandemic has disrupted the economic and social infrastructure of countries across the world. Academic institutions have faced the burnt particularly affecting both the physical and psychological well-being of school-going children. As the world is trying to recover from the catastrophic after-effects of the pandemic, lessons learned from this global health crisis remain relevant.

Our IoT based 'SecureTrack' platform provides an innovative solution for contact tracing and monitoring of infectious diseases that can be applied well beyond the COVID-19 outbreak. Its wearable design and data privacy make it a valuable tool for academic institutions looking to reopen quickly while preventing the spread of infectious diseases. The platform's scalability and ease of use also make it an attractive option for large clusters of organizations in need of contact tracing solutions. A prototype of our device was developed on the MICAz platform and successfully verified to validate our approach.

## References

[1] C. for Systems Science, E. C. at the Johns Hopkins University, Coronavirus 370 covid-19 (2019-ncov), https://gisanddata.maps.arcgis.com/apps/ opsdashboard/index.html#/bda7594740fd40299423467b48e9ecf6.

[2] M. S. H. J. B. S. Dae-Gyun Ahn, Hye-Jin Shin, Current status of epidemiology, diagnosis, therapeutics, and vaccines for novel coronavirus disease 2019 (covid-19), Journal of Microbiology and Biotechnology 30 (3) 375 (2020) 313–324, http://jmb.or.kr/journal/view.html?doi=10.4014/ jmb.2003.03011. doi:10.4014/jmb.2003.03011.

[3] N. Lurie, M. Saville, R. Hatchett, J. Halton, Developing covid-19 vaccines at pandemic speed, New England Journal of Medicine 382 (21) (2020) 1969–1973, https://doi.org/10.1056/NEJMp2005630. arXiv:https:// 380 doi.org/10.1056/NEJMp2005630, doi:10.1056/NEJMp2005630.

[4] C. Liu, Q. Zhou, Y. Li, L. V. Garner, S. P. Watkins, L. J. Carter, J. Smoot, A. C. Gregg, A. D. Daniels, S. Jervey, D. Albaiu, Research and development on therapeutic agents and vaccines for covid-19 and related human coronavirus diseases, ACS Central Science 6 (3) (2020) 315–331, https://doi. 385 org/10.1021/acscentsci.0c00272. doi:10.1021/acscentsci.0c00272.

[5] Provisional covid-19 deaths: Focus on ages 0-18 years — data — centers for disease control and prevention, https://data.cdc.gov/NCHS/ Provisional-COVID-19-Deaths-Focus-on-Ages-0-18-Yea/nr4s-juj3.

[6] Pastorino R, Pezzullo AM, Villani L, Causio FA, Axfors C, Contopoulos390 Ioannidis DG, Boccia S, Ioannidis JPA. Change in age distribution of COVID-19 deaths with the introduction of COVID-19 vaccination. Environ Res. 2022 Mar;204(Pt C):112342. doi: 10.1016/j.envres.2021.112342. Epub 2021 Nov 5. PMID: 34748775; PMCID: PMC8570444.

[7] H. Lau, V. Khosrawipour, P. Kocbach, A. Mikolajczyk, J. Schubert, J. Ba395 nia, T. Khosrawipour, The positive impact of lockdown in Wuhan on containing the COVID-19 outbreak in China, Journal of Travel Medicine 27 (3), taaa037. arXiv:https://academic.oup.com/jtm/article-pdf/ 27/3/taaa037/33226344/taaa037.pdf, doi:10.1093/jtm/taaa037. URL https://doi.org/10.1093/jtm/taaa037 400

[8] Colbourn T. COVID-19: extending or relaxing distancing control measures. Lancet Public Health. 2020;5(5):e236-e237. doi:10.1016/S2468- 2667(20)30072-4.

[9] Rossi R, Socci V, Talevi D et al. COVID-19 Pandemic and Lockdown Measures Impact on Mental Health Among the General Popu405 lation in Italy. Front Psychiatry. 2020;11:790. Published 2020 Aug 7. doi:10.3389/fpsyt.2020.00790.

[10] R. Rossi, V. Socci, D. Talevi, S. Mensi, C. Niolu, F. Pacitti, A. Di Marco, A. Rossi, A. Siracusano, G. Di Lorenzo, Covid-19 pandemic and lockdown measures impact on mental health among the general population in italy, 410 Frontiers in Psychiatry 11 (2020) 790. doi:10.3389/fpsyt.2020.00790. URL https://www.frontiersin.org/article/10.3389/fpsyt.2020. 00790

[11] U. Rehman, M. G. Shahnawaz, N. H. Khan, K. D. Kharshiing, M. Khursheed, K. Gupta, D. Kashyap, R. Uniyal, Depression, anxiety and stress 415 among indians in times of covid-19 lockdown, Community Mental Health Journaldoi:10.1007/s10597-020-00664-x. URL https://doi.org/10.1007/s10597-020-00664-x.

[12] P. Majumdar, A. Biswas, S. Sahu, Covid-19 pandemic and lockdown: cause 20 of sleep disruption, depression, somatic pain, and increased screen expo420 sure of office workers and students of india, Chronobiology International 0 (0) (2020) 1–10, pMID: 32660352. arXiv:https://doi.org/10.1080/ 07420528.2020.1786107, doi:10.1080/07420528.2020.1786107. URL https://doi.org/10.1080/07420528.2020.1786107.

[13] P. Odriozola-Gonz´alez, Alvaro Planchuelo-G´omez, M. J. Irurtia, R. de ´ 425 Luis-Garc´ıa, Psychological effects of the covid-19 outbreak and lockdown among students and workers of a spanish university, Psychiatry Research 290 (2020) 113108, "http://www.sciencedirect.com/science/ article/pii/S0165178120313147". doi:https://doi.org/10.1016/j. psychres.2020.113108. 430

[14] Gualano, M.R., Lo Moro, G., Voglino, G., Bert, F., Siliquini, R.. Effects of Covid-19 Lockdown on Mental Health and Sleep Disturbances in Italy. Int. J. Environ. Res. Public Health 2020, 17, 4779.

[15] Thakur V, Jain A. COVID 2019-suicides: A global psychological pandemic. Brain Behav Immun. 2020 Aug;88:952-953. doi: 10.1016/j.bbi.2020.04.062. 435 Epub 2020 Apr 23. PMID: 32335196; PMCID: PMC7177120.
[16] Covid-19 asymptomatic, https://jamanetwork.com/journals/ jamanetworkopen/fullarticle/2787098.

[17] I. F. Miller, A. D. Becker, B. T. Grenfell, C. J. E. Metcalf, Disease and healthcare burden of covid-19 in the united states, Nature Medicine 26 (8) 440 (2020) 1212–1217, https://doi.org/10.1038/s41591-020-0952-y. doi: 10.1038/s41591-020-0952-y.

[18] Cdc activities and initiatives supporting the covid-19 response and the president's plan for opening america up again, https://www.cdc.gov/coronavirus/2019-ncov/downloads/php/ 445 CDC-Activities-Initiatives-for-COVID-19-Response.pdf#page=53

[19] A. D. Paltiel, A. Zheng, R. P. Walensky, Assessment of SARS-CoV-2 Screening Strategies to Permit the Safe Reopening of College Campuses in the United States, JAMA Network Open 3 (7) (2020) e2016818–e2016818, https://doi.org/10.1001/jamanetworkopen.2020.16818. arXiv: 450 https://jamanetwork.com/journals/jamanetworkopen/articlepdf/ 2768923/paltiel\_2020\_oi\_200614\_1597251392.14363.pdf, doi:10.1001/jamanetworkopen.2020.16818.

[20] R. M. Viner, C. Bonell, L. Drake, D. Jourdan, N. Davies, V. Baltag, J. Jerrim, J. Proimos, A. Darzi, Reopening schools during the covid-19 pan455 demic: governments must balance the uncertainty and risks of reopening schools against the clear harms associated with prolonged closure, Archives of Disease in Childhoodhttps://adc.bmj.com/content/early/2020/ 08/02/archdischild-2020-319963. arXiv:https://adc.bmj.com/ content/early/2020/08/02/archdischild-2020-319963.full.pdf, 460 doi:10.1136/archdischild-2020-319963.

[21] Case investigation and contact tracing : Part of a multipronged approach to fight the covid-19 pandemic, https://www.cdc.gov/ coronavirus/2019-ncov/php/principles-contact-tracing.html#: ~:text=Support%20isolation%20of%20those%20who,symptoms%20to%465 20testing%20and%20care.

[22] Mei, X., Lee, H., Diao, K. et al. Artificial intelligence–enabled rapid diagnosis of patients with COVID-19. Nat Med 26, 1224–1228 (2020). https://doi.org/10.1038/s41591-020-0931-3.

[23] "V. Chamola, V. Hassija, V. Gupta and M. Guizani, "A Comprehensive 470 Review of the COVID-19 Pandemic and the Role of IoT, Drones, AI, Blockchain, and 5G in Managing its Impact," in IEEE Access, vol. 8, pp. 90225-90265, 2020, doi: 10.1109/ACCESS.2020.2992341.".

[24] "Otoom M, Otoum N, Alzubaidi MA, Etoom Y, Banihani R, An IoTbased Framework for Early Identification and Monitoring of COVID-19 Cases, Biomedical Signal Processing and Control (2020), doi: https://doi.org/10.1016/j.bspc.2020.102149".

[25] "R. Abbas and K. Michael, "COVID-19 Contact Trace App Deployments: Learnings From Australia and Singapore," in IEEE Consumer Electronics Magazine, vol. 9, no. 5, pp. 65-70, 1 Sept. 2020, doi: 480 10.1109/MCE.2020.3002490.".

[26] "Hyunghoon Cho, Daphne Ippolito and Yun William Yu, "Contact Tracing Mobile Apps for COVID-19: Privacy Considerations and Related Tradeoffs, arXiv:2003.11511v2 [cs.CR] 30 Mar 2020.".

[27] "Ramaci, T.; Barattucci, M.; Ledda, C.; Rapisarda, V. Social Stigma dur485 ing COVID-19 and its Impact on HCWs Outcomes. Sustainability 2020, 12, 3834.".

[28] "Bagcchi S. Stigma during the COVID-19 pandemic. Lancet Infect Dis. 2020 Jul;20(7):782. doi: 10.1016/S1473-3099(20)30498-9. PMID: 32592670; PMCID: PMC7314449.". 490

[29] "Logie, C.H., Turan, J.M. How Do We Balance Tensions Between COVID-19 Public Health Responses and Stigma Mitigation? Learning from HIV Research. AIDS Behav 24, 2003–2006 (2020). https://doi.org/10.1007/s10461-020-02856-8".

[30] M. E. Berglund, J. Duvall, L. E. Dunne, A survey of the historical scope 495 and current trends of wearable technology applications, in: Proceedings of the 2016 ACM International Symposium on Wearable Computers, ISWC '16, Association for Computing Machinery, New York, NY, USA, 2016, p. 40–43. doi:10.1145/2971763.2971796.

[31] A. Godfrey, V. Hetherington, H. Shum, P. Bonato, N. Lovell, 500 S. Stuart, From a to z: Wearable technology explained, Maturitas 113 (2018) 40 – 47, "http://www.sciencedirect.com/science/ 23 article/pii/S0378512218302330". doi:https://doi.org/10.1016/j. maturitas.2018.04.012.

[32] S. Sengupta, J. Kim, Seong Dae Kim, Applying triz and bass model to 505 forecast fitness tracking devices technology, in: 2015 Portland International Conference on Management of Engineering and Technology (PICMET), 2015, pp. 2177–2186.

[33] J. Casselman, N. Onopa, L. Khansa, Wearable healthcare: Lessons from the past and a peek into the future, Telematics and Informatics 510 34 (7) (2017) 1011 – 1023,

"http://www.sciencedirect.com/science/ article/pii/S0736585316307274". doi:https://doi.org/10.1016/j. tele.2017.04.011.

[34] A. Marakhimov, J. Joo, Consumer adaptation and infusion of wearable devices for healthcare, Computers in Human Behavior 76 (2017) 515 135 – 148, "http://www.sciencedirect.com/science/article/pii/ S0747563217304284". doi:https://doi.org/10.1016/j.chb.2017.07. 016.

[35] A. A. Mohammed, P. Mohsen, Market-driven management of startups: The case of wearable technology ahead-of-print (ahead-of-print) 520 (2020) 2210–8327 A1 – Dehghani Milad, https://doi.org/10.1016/j. aci.2018.11.002. doi:10.1016/j.aci.2018.11.002.

[36] C. Glaros, D. I. Fotiadis, Wearable devices in healthcare (2005) 237– 264"https://doi.org/10.1007/11311966_8". doi:10.1007/11311966_ 8. 525

[37] K. Hung, Y. T. Zhang, B. Tai, Wearable medical devices for tele-home healthcare, in: The 26th Annual International Conference of the IEEE Engineering in Medicine and Biology Society, Vol. 2, 2004, pp. 5384–5387.

[38] How wearable tech is enhancing our everyday lives, https://enlightened-digital.com/ 24 530 how-wearable-tech-is-enhancing-our-everyday-lives/#:~: text=Wearable%20devices%20available%20today%20can,and% 20better%20understand%20our%20health.

[39] P. P. Ray, Home health hub internet of things (h3iot): An architectural framework for monitoring health of elderly people, in: 2014 International 535 Conference on Science Engineering and Management Research (ICSEMR), 2014, pp. 1–3.

[40] J. Mohammed, C. Lung, A. Ocneanu, A. Thakral, C. Jones, A. Adler, Internet of things: Remote patient monitoring using web services and cloud computing, in: 2014 IEEE International Conference on Internet of Things 540 (iThings), and IEEE Green Computing and Communications (GreenCom) and IEEE Cyber, Physical and Social Computing (CPSCom), 2014, pp. 256–263.

[41] G. Yang, L. Xie, M. M¨antysalo, X. Zhou, Z. Pang, L. D. Xu, S. KaoWalter, Q. Chen, L. Zheng, A health-iot platform based on the integration 545 of intelligent packaging, unobtrusive bio-sensor, and intelligent medicine box, IEEE Transactions on Industrial Informatics 10 (4) (2014) 2180–2191.

[42] A. Mdhaffar, T. Chaari, K. Larbi, M. Jmaiel, B. Freisleben, Iot-based health monitoring via lorawan, in: IEEE EUROCON 2017 -17th International Conference on Smart Technologies, 2017, pp. 519–524. 550

[43] Xbee 3 series datasheet, https://www.digi.com/resources/library/ data-sheets/ds_xbee3-802-15-4.

[44] L. Panxing, W. Tong, A method for adjusting transmit power of zigbee network node based on rssi, in: 2015 IEEE International Conference on Signal Processing, Communications and Computing (ICSPCC), 2015, pp. 555 1–4.

[45] Honor band 5 specifications, https://www.hihonor.com/global/ products/wearables/honorband5/.

[46] Xiaomi band 4 specifications, https://www.mi.com/uk/ mi-smart-band-4/. 560 [47] Micaz datasheet, http://www.cmt-gmbh.de/Produkte/ WirelessSensorNetworks/Datenblaetter/MICAz_Datasheet.pdf.

[48] Path loss exponent for various environments, https://personal. utdallas.edu/~torlak/courses/ee4367/lectures/lectureradio.pdf.